\documentclass{ws-p8-50x6-00_changed_for_losalamospreprints}

\begin{document}






\noindent
University of Bern, Laboratory for High Energy Physics
\\
preprint BUHE-9906
\\

\vspace{0.5cm}

\begin{center}
{\large \bf CRYOGENIC CALORIMETERS IN ASTRO AND PARTICLE PHYSICS}
\\
\vspace{0.5cm}
\vspace{0.5cm}

{K.PRETZL}
\vspace{0.2cm}

{ \it Laboratory for High Energy Physics of the University Bern, Sidlerstr. 5 \\
CH 3012 Bern, Switzerland
E-mail:Pretzl@lhep.unibe.ch}

\end{center}
\vspace{0.5cm}
\vspace{0.5cm}

\abstracts{The development of cryogenic calorimeters was originally motivated by the fact 
that very low energy thresholds and excellent energy resolutions can be achieved by 
these devices. Cryogenic devices are widely used in double beta decay experiments, 
in cosmological dark matter searches, in x-ray detection of galactic and extragalactic 
objects as well as in cosmic background radiation experiments. An overview of the latest 
developments is given.}

\section{Introduction}
Cryogenic calorimeters open up the possibility of measuring processes with energy
transfers as low as eV with very high accuracy. Their development has recently
been motivated by the quest for the dark matter in the universe and for
 the missing 
neutrinos from the sun. Other research areas have also benefitted from these
developments, such as the double $\beta $-decay experiments, neutrino mass experiments,
x-ray spectroscopy in astrophysics, single photon counting spectrometry , 
mass spectrometry of large molecules (DNA sequencing) as
well as x-ray microanalysis for industrial applications.

\section{First Ideas and Attempts}
In 1935 F.Simon \cite{si} suggested measuring the energy deposited by radioactivity 
with cryogenic calorimeters. Later in 1949 D.H.Andrews, R.D.Fowler and
M.C.Williams \cite{an} reported the detection of individual $\alpha$-particles using a 
superconducting strip. G.H.Wood and B.L.White \cite{wo} detected $\alpha $-particles with
a superconducting tunnel junction (STJ) detector. H.Bernas et al. \cite{be} used
superheated superconducting granules (SSG) for beta radiation. A.K.Drukier and
C.Vallette detected charged particles with SSG \cite{dr}. T.Niinikoski and
F.Udo \cite{ni} proposed cryogenic calorimeters for the detection of neutrinos in
1974. E.Fiorini and T.O.Niinikoski \cite{fi} explored in 1983 the possibility of 
low temperature calorimetry to improve the limits on processes such as neutrinoless
double-beta decay. A.Drukier and L.Stodolsky \cite{dru} suggested 1984 the use
of SSG detectors for neutrino-physics and astrophysics experiments. With all these
interesting first ideas and experimental attempts in mind, a first workshop (LTD1)
\cite{pre} on low temperature detectors was organized in 1987 ­ at Ringberg-Kastell
on Lake Tegernsee in southern Bavaria. Many of these ideas have been shown to work
and the field has been successfully growing and reaching many areas of research.
Subsequent workshops were held in Annecy (France) LTD2 \cite{go}, the Gran Sasso Laboratory 
(Italy)
LTD3 \cite{br}, Oxford (Great Britain) LTD4 \cite{bo}, Berkeley (USA) LTD5 \cite{la}, Beatenberg 
(Switzerland) LTD6
\cite{ot} and Munich (Germany) LTD7 \cite{co}. The eighth in this series of workshops LTD8
will be held in Dalfsen (Netherlands)(15-20 August 1999). Much of the original work in
this field can be found in the proceedings of these workshops as well as in \cite{ba}.
There are also excellent review articles on the subject \cite{bar,tw,kra,boot}.

\section{Why Cryogenic Detectors?}
Most calorimeters used in high energy physics measure the energy loss of a particle
in form of ionization or scintillation light. In contrast, cryogenic calorimeters are able to 
measure the total deposited energy in form of ionization and heat. This feature makes them
very effective in detecting very small energy deposits with high resolution. A small energy loss $\Delta $E of 
a particle can lead to an appreciable temperature increase $\Delta $T in the calorimeter

\begin{center}
\begin{equation}
{\Large \Delta T=\frac{\Delta E}{C_{tot}}}
\label{eq:e1}
\end{equation}
\end{center}
\noindent
provided the  heat capacity $C_{tot}=cV$ of a calorimeter 
with a given volume V is small. A condition which can be reached at low temperatures
due to the rapid decrease of the specific heat c with temperature\\

\noindent
for a dielectric crystal 
\begin{equation}
{\Large 
c = \beta \  [\frac {T} {\Theta_D} ]^3}
\label{eq:e2}
\end{equation}
\noindent
for a normal conductor
\begin{equation}
{\Large 
c = \beta \  [ \frac {T} {\Theta_D} ]^3 + \gamma \   T}
\label {eq:e3}
\end{equation}
\noindent
for a superconductor
\begin{equation}
{\Large 
c = \beta \ [ \frac {T} {\Theta_D} ]^3 + \alpha \  e^{\Delta/kT}}
\label {eq:e4}
\end{equation}
\noindent								
where $\alpha,\beta,\gamma$  are material dependent parameters, $\Theta_D $ is the Debye temperature 
and $\Delta$ is the energy 
gap of the superconductor with a typical value of the order of 1meV.

In Fig.1 a typical cryogenic calorimeter is shown. It consists of an absorber
with heat capacity C, a thermometer and a thermal link with heat conductance g to
a heat reservoir with constant bath temperature T$_B$. Particles interacting 
in the absorber cause a change in the resistance of the 
themometer. This is measured by observing a voltage drop accross it when passing a current I
through the thermometer.

\begin{figure}[t]
\epsfxsize=20pc 
\epsfysize=20pc 
\hspace*{1.5cm}
\epsfbox{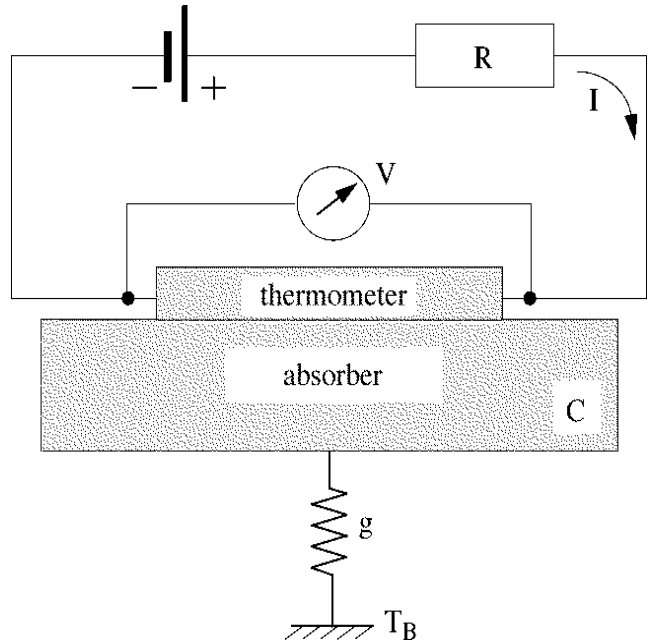} 
\caption{
The principle of a cryogenic calorimeter is shown.
  \label{fig:figure1}}
\end{figure}

Cryogenic calorimeters can be made from many different materials including superconductors,
a feature which turns out to be very useful for many applications. They can be used as targets 
and detectors at the same time. Due to the very small energy quanta involved they reach 
much higher energy resolutions than conventional devices. For example, it takes only
of the order of 1meV to break a Cooper pair in a superconductor whereas a few eV are needed 
to create an electron-hole pair in a solid-state device.

\section{Energy Resolution}
The mean energy fluctuation in an absorber with a thermal link to the heat sink
is:

\begin{center}
\begin{equation}
\large \Delta E_{FWHM}=2.35 \xi \sqrt{kT^{2}C}
\end{equation}
\end{center}
\noindent

It is independent of the absorbed energy and the thermal
conductance of the link. $\xi $ is a parameter which depends on the sensitivity and
noise characteristics of the thermometer and can have values between 1.2 and 2.0.

The energy resolution of a semiconductor device is: 
\begin{center}
\begin{equation}
\large \Delta E_{FWHM}=2.35 \sqrt{w F E} \sim 110eV
\end{equation}
\end{center}
\noindent
for a 6keV x-ray, where w is the average energy necessary to produce an electron hole
pair. It has a typical value of w$\approx$3eV. The Fano factor is F = 0.12 for Silicon.

The use of superconductors as cryogenic particle detectors was motivated by the
small binding energy 2$\Delta\approx$1meV of the Cooper pairs. A particle traversing a
superconductor produces quasiparticles (by breaking Cooper pairs) and phonons. As long as
the energy of the quasiparticles and phonons is higher than 2$\Delta$, they break up
more Cooper pairs and continue to produce quasiparticles until their energy falls
below the threshold of 2$\Delta$. Thus, compared to a semiconductor, several orders of
magnitude more free charges are produced leading to a much higher intrinsic energy resolution.

For a superconducting
cryogenic calorimeter with a Cooper pair binding energy of 2$\Delta \approx $1meV , the best
obtainable energy resolution is

\begin{center}
\begin{equation}
\large \Delta E_{FWHM}=2.35 \sqrt{2\Delta F E} \sim 3eV
\end{equation}
\end{center}
\noindent
for a 6keV x-ray assuming F=0.2 for most superconductors.

In Fig.2 x-ray spectra obtained with a state of the art Si(Li) solid-state device and 
a cryogenic microcalorimeter using a HgCdTe absorber are compared. The microcalorimeter 
was a rather early version developed by the Wisconsin, NASA Goddard group and had an 
energy resolution of 35eV. In the meantime this group \cite{camm}and others\cite{al,hil}
have succeeded in building microcalorimeters with energy resolutions between 3 and 5eV. Such 
calorimeters have typically a surface area of $\leq 1mm^2$ and a
thickness of several $\mu $m.

\begin{figure}[t]
\epsfxsize=22pc 
\epsfysize=30pc 
\hspace*{1.5cm}
\epsfbox{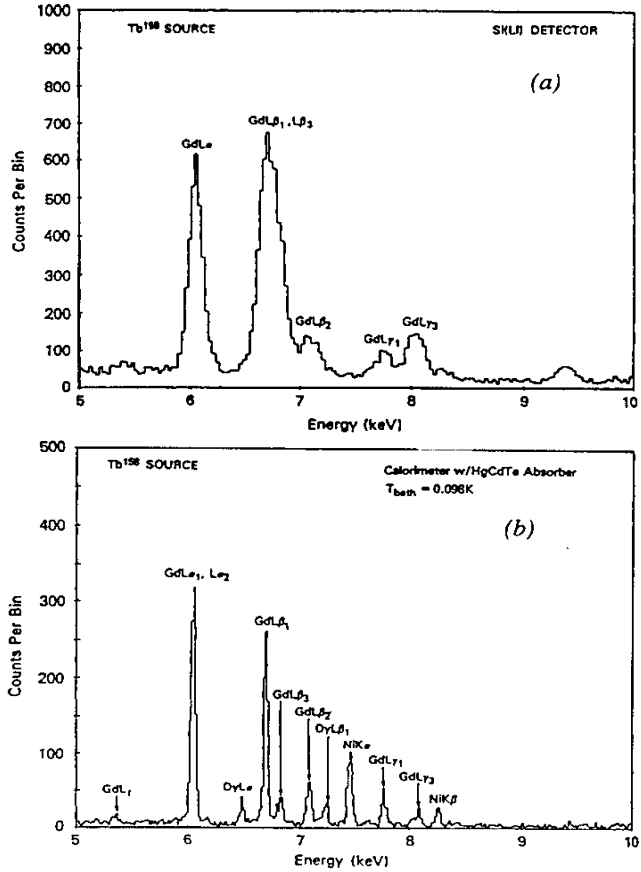} 
\caption{
X-ray spectra obtained (a) with a Si(Li) solid state
detector and (b) with a HgCdTe micro-calorimeter 
(from the Wisconsin/Goddard group) are compared.
\label{fig:figure2}}
\end{figure}

\section{Phonon Sensors}
Phonons produced by a particle interaction in an absorber are far from thermal
equilibrium. They must decay to lower energy phonons and become thermalized
before the temperature rise $\Delta $T can be measured. Since thermalisation 
results in a rather long decay time (order of msec) of the signal pulse, a cryogenic
phonon detector cannot tolerate counting rates much higher than a few Hz.

The most commonly used phonon sensors are:
Semiconducting thermistors and superconducting transition edge sensors (TES).

\subsection{Semiconducting Thermistors}\label{subsec:semi}
A thermistor is a heavily doped semiconductor
slightly below the metal insulator transition. Good uniformity of doping
concentrations can be achieved either with ion implantation or with neutron
transmutation doping (NTD). In the latter case, 
thermal neutrons from a reactor are
captured by nuclei which transform into isotopes. These can then be the donors or
acceptors for the semiconductor. NTD Ge thermistors are frequently used, because of their 
reproducibility and their uniformity in doping density. Furthermore they are easy to 
handle and commercially available. However, they have the disadvantage of being 
thermal phonon sensors with intrinsically slow response signals and of having to deal 
with Joule heating.

\subsection{Superconducting Transition Edge Sensors (TES)}\label{subsec:super}

An alternative thermometer is a
thin superconducting strip which is operated close to its superconducting to
normal phase transition temperature Tc. The operating temperature is chosen so
that it is sensitive in the region of the temperature versus resistance diagram as
shown in Fig.3a. The voltage drop across the strip, i.e. the signal pulse height,
depends on the magnitude of the constant current  through the strip and on the
change of the resistance $\Delta $R. However, the current should be kept as small as
possible in order to reduce the self-heating effects. The Munich group \cite{colli} has
developed a transition edge sensor made out of tungsten (W) with a transition temperature of
15 mK. They used low impedance sensors with normal conducting resistances between
10 and 100m$\Omega$. A constant current is fed through the readout circuit as shown in
Fig.3b. An increase in R(T) due to a temperature rise forces more current through
the parallel branch of the circuit, inducing a magnetic flux change in L which is
measured with high sensitivity by a SQUID. However, in this mode the temperature of 
the detector has to be stabilized
with very high precision in order to achieve a good energy resolution. Recently
K.D.Irwin \cite{irw} has developed an interesting auto-biasing electrothermal
feedback system, which works as a thermal equivalent to an operation amplifier.
It keeps the temperature of the superconducting strip at a constant value within
its transition region. When operating the transition edge sensor in a voltage biased
mode (V$_B$), a temperature rise in the sensor causes an increase in its resistance
and a corresponding decrease in the current, which results in a decrease of the
Joule heating (V$_B\cdot\Delta $I). The feedback uses the decrease of the Joule 
heating to bring
the temperature of the strip back to the constant operating value. Thus the
device is self calibrating. The deposited energy in the absorber is given by 
E =-V$_B\int\Delta $I(t)dt.

\begin{figure}[t]
\epsfxsize=30pc 
\epsfysize=18pc 
\epsfbox{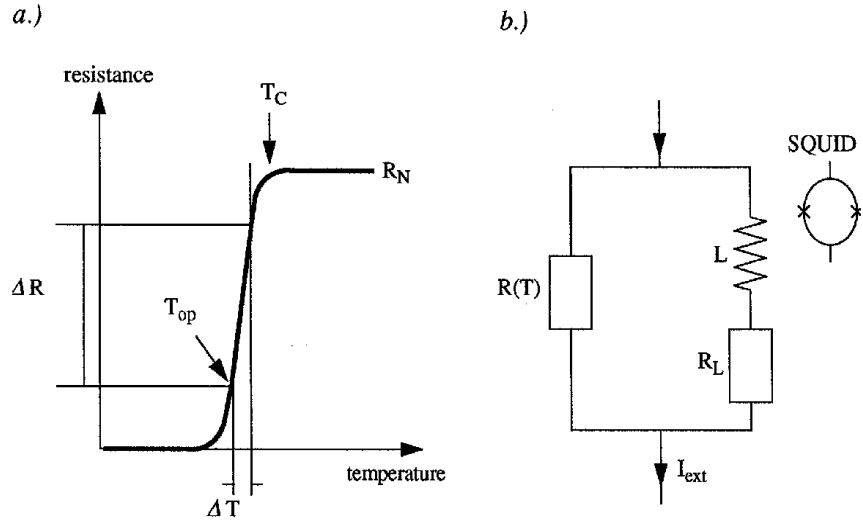} 
\caption{
(a) The temperature versus resistance diagram of a superconducting strip
close to the transition temperature T$_c$ is shown.
(b) The dc-SQUID readout of a transition edge sensor is shown.
 \label{fig:figure3}}
\end{figure}

Other transition edge sensors, made from 
proximity bilayers such as Al/Ag,Al/Cu,Ir/Au,Mo/Au,and Mo/Cu, have been developed to cover
also higher transition temperatures in the range between 15 and 150mK.  

TES is  sensitive to non-thermal phonons with energies well above
2$\Delta $. While losing energy these phonons produce quasiparticles before they thermalize. 
Since this process is very much faster than thermalization, signals of 
the order of $\mu $s can be achieved, enhancing considerably  the counting rate capability 
of these devices as compared to thermal phonon sensors. Due to this circumstance and the selfcalibrating 
effect of the electrothermal feedback system, 
TES are now among the most frequently used 
devices in calorimetric measurements.

\section{Quasiparticle Detection}

Quasiparticles produced by the absorption of X-rays or the energy loss of a transient particle 
in a superconducting absorber can be measured with a Superconducting Tunnel Junction (STJ). When
biasing the STJ at a suitable voltage the tunneling current through the junction is proportinal 
to the excess number of quasiparticles produced.
 Arrays of STJ¹s are also used to measure high energy, non-thermal (ballistic) phonons produced in
either a dielectric or superconducting absorber. An excellent educational description of STJ¹s
can be found in \cite{baro}.

\subsection{Superconducting Tunnel Junctions (STJ)}\label{subsec:tunnel}

A STJ consists of two superconducting films with a thickness of typically a few
nm separated by a thin, 1-2 nm thick, tunnel barrier, which is usually the
oxide of one of the superconductors. Typical junction areas are of the order of
100 x 100$\mu $m$^{2}$ . As a quasiparticle detector, the STJ is operated with a bias
voltage which is usually set to be less than $\Delta $/e. A magnetic field of about 100
Gauss is applied in the plane of the junction in order to suppress the Josephson
supercurrent. A change in the number of quasiparticles will lead to a tunneling
current which is proportional to the number of quasiparticles produced in the
absorber . Since the Poisson statistical fluctuations of the quasiparticle
density in the superconductor are very small, the two most important parameters
driving the energy resolution of the STJ are the tunneling rate $\tau _{tunnel}^{-1} $   and the thermal
recombination rate $\tau _{re}^{-1} $. The temperature dependence of the thermal recombination rate
is given by

\begin{center}
\begin{equation}
\large \tau ^{-1}_{re}(T)=\tau ^{-1}_{0} \sqrt{\pi} \left(\frac{2\Delta}{kT_{c}}\right)^{5/2} \sqrt{\frac{T}{T_{c}}} \exp{(-\frac{\Delta}{kT})} 
\end{equation}
\end{center}
\noindent
where $\tau _0 $ is the characteristic time of a superconductor. It has the
values $\tau _0 $ = 2.3ns for Sn, $\tau _0 $ = 438ns for Al and $\tau _0 $ = 0.15ns for Nb \cite{ka}.

The recombination rate $\tau _{re}^{-1} $ can be minimized when operating the detector at
sufficiently low temperatures, typically at 0.1 Tc, where the number of thermally
excited quasiparticles is very small. The tunneling time is given by
\begin{center}
\begin{equation}
\large \tau ^{-1}_{tunnel}=R_{norm} \cdot e^{2} \cdot N_{0} \cdot A \cdot d
\end{equation}
\end{center}
\noindent
 where R$_{norm} $ is the normal conducting resistance of the junction, N$_0 $ 
 is the density of states
of one spin at the Fermi energy, A is the junction overlap area and d the
thickness of the corresponding film. In practice, the tunneling time has to be
shorter than the quasiparticle lifetime. For 100 x 100$\mu $m$^{2}$  junctions a tunneling
time of the order of 1$\mu $s  has been obtained. In order to achieve even shorter
tunneling times, one would have to try to further reduce R$_{norm} $. However, there is a
fabricational limit avoiding micro-shorts in the insulator between the
superconducting films. Quasiparticles are lost for detection when they produce
sub-gap phonons with energies below 2$\Delta $ (which do not contribute to the detector
signal) or when they diffuse out of the overlap region of the junction films
into the current leads instead of crossing the junction gap. N.Booth \cite{boot}
proposed a scheme (Fig.4) which allows to recover some of these losses by the use
of quasiparticle trapping and in some cases even quasiparticle multiplication.
The scheme has been shown to work successfully. The quasiparticles produced in
the superconducting absorber S$_1 $ diffuse to the superconducting film S$_2 $ of the STJ with a smaller
gap energy $\Delta _2 $. By falling in that trap they relax to smaller energies by emitting
phonons, which could produce additional quasiparticles in the film S$_2 $
(quasiparticle multiplication) if their energy is larger than $\Delta _2 $. The relaxed
quasiparticles cannot diffuse back into the superconductor S$_1 $ because of their
lower energy. They will eventually tunnel through the STJ, contributing to the
signal. In order for quasiparticle trapping to be effective, superconducting
absorber materials with long quasiparticle lifetimes have to be selected (for
example Al).

\begin{figure}[t]
\epsfxsize=20pc 
\epsfysize=20pc 
\hspace*{1.5cm}
\epsfbox{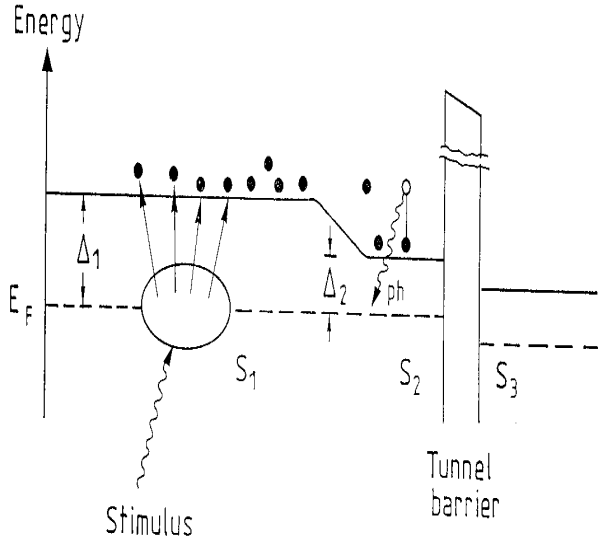} 
\caption{
The schematic of quasiparticle trapping is shown.
See text for explanation.
  \label{fig:figure4}}
\end{figure}

Because of their limited size, STJ¹s have so far not been used in connection with
large volume calorimeters. This may, however, change in the near future because
large arrays of STJ¹s can be produced by now. The Naples group \cite{cris} has recently
produced an array of circular shaped STJ¹s.

STJ¹s are commonly used for high
resolution x-ray detection. High energy resolutions of 30 eV for 6 keV x-rays have
been obtained by the ESTEC \cite{peac} and Livermoore \cite{labov} groups. As described below,
the ESTEC group achieved single photon counting with niobium and tantal STJ¹s.

\subsection{Superheated Superconducting Granules (SSG)}\label{subsec:granules}

Superheated superconducting granules (SSG) have been developed for x-ray imaging,
transition radiation, dark matter and solar neutrino detection. A SSG detector
consists of billions of small grains (with a diameter of 30$\mu$m for example), diluted in
a dielectric material (e.g. Teflon) with a volume filling factor of typically 10$\%$. The
detector is operated in an external magnetic field. Metastable type-1  superconductors 
(e.g. Sn. Zn, Al, Ta) are used, since their phase transitions from the metastable 
superconducting state to the normal 
state are sudden (in the order of 100ns) allowing for a fast time correlation between SSG 
signals and those of other detectors.
 The SSG detector is a threshold detector operated 
typically at a temperature of T=0.1 T$_c$. 
Its energy threshold is adjustable by setting the external magnetic field at a certain 
value just below the phase transition border.  
  Particles interacting in a granule produce 
quasiparticles. While spreading over the volume of the granule the quasiparticles are 
losing energy via electron-phonon
interactions,
 thereby globally heating the granule up to a point where it may undergo  
a sudden phase transition (granule flip).  The phase transition of individual grains in a large
sample  can be detected by a pickup loop, which measures the magnetic
flux change due to the disappearance of the Meissner-Ochsenfeld effect. The Bern group has developed a
 readout system which allows to detect single grain flips ( grains with 30$\mu$m diameter) 
 in a pickup coil of 2cm diameter 
and 16cm length. As described below, 50 such pickup coils are needed for a 1kg SSG dark 
matter detector.
In parallel, a  more sensitive dc-SQUID (Superconducting Quantum Interference Devices) 
 system is under development to increase 
the sensitivity of the  detector by allowing to read smaller size granules\cite{vande}.

Small spherical grains can be produced at low cost by industry. Since after
fabrication the grains are not of a uniform diameter, they  have to be sieved
 to select the desired size. A grain size selection
within $\pm$2$\mu$m was achieved.

However, it was discovered that industrially produced grains do not exhibit a
sharp phase transition boundary. Their phase transition is smeared by about 20$\%$,
leading to an equivalent smearing of the energy threshold. This phase transition smearing 
could only very recently be reduced by
laser treatment and fast cooling of the grains  by as much as an order of magnitude
\cite{cal}. Similar results were also obtained by
producing small cylindrical Sn structures on a substrate by evaporation techniques. Both methods
 seem to allow the production of large quantities of granules for a massive
SSG dark matter detector.

A superconducting superheated granule detector offers several unique features:
a) The large list of suitable type-1 superconductor materials allows to optimize SSG
for specific applications.
b) Very low energy thresholds (eV) can be achieved.
c) The pickup loop readout system does not 
dissipate any energy into grains. Therefore the sensitivity of SSG is essentially 
determined by the grain size and the specific heat of the grain material.
d) The sudden phase transitions are beneficial for coincident timing with other signals. 

Generally speaking, SSG detectors are among the most sensitive devices to detect very low energy transfers, i.e.
nuclear recoils. A detailed description of SSG can be found in \cite{pr,pret}.

\section{Cryogenic Detectors and the Quest for Dark Matter in the Universe}

There is evidence that a significant amount of the dark matter in the Universe is
of exotic, i.e. non-baryonic, nature. Prime candidates are massive neutrinos,
axions and also neutralinos, the lightest SUSY particles predicted by Super
Symmetry. But there is plenty of room for other species, which are generally
called WIMP¹s (weakly interacting massive particles).

WIMP¹s can be detected by measuring the nuclear recoil energy in elastic
WIMP-nucleus scattering. Depending on the mass of the WIMP and the mass of 
the detector nucleus, the average recoil energy can vary between eV and keV.
The main advantage of cryogenic devices for the  WIMP search is their
effectiveness in detecting very low energy nuclear recoils and the possibility to
use a large variety of detector materials. The
first generation WIMP experiments with cryogenic detectors are being done with 
absorber masses of about 1kg. Larger detector masses of 100 kg are planned
for the future. A list of cryogenic WIMP detectors in preparation or already in
operation is given in Table 1.

\begin{table}[t]
\caption{WIMP searches with cryogenic calorimeters.}
\begin{center}
\footnotesize
\begin{tabular}{|l|l|l|}
\hline
\raisebox{0pt}[13pt][7pt]{\large \bf Experiment} 
& \raisebox{0pt}[13pt][7pt]{\large \bf Location} 
& \raisebox{0pt}[13pt][7pt]{\large \bf Absorber} \\
\hline
\raisebox{0pt}[13pt][7pt]{CDMS}   & \raisebox{0pt}[13pt][7pt]{Stanford, U.S.A.}    &  \raisebox{0pt}[13pt][7pt]{Germanium, Silicium} \\
\hline
\raisebox{0pt}[13pt][7pt]{CRESST} & \raisebox{0pt}[13pt][7pt]{Gran Sasso, Italy}   &  \raisebox{0pt}[13pt][7pt]{Sapphire } \\
\hline
\raisebox{0pt}[13pt][7pt]{EDELWEISS}  &  \raisebox{0pt}[13pt][7pt]{Frejus, France}   & \raisebox{0pt}[13pt][7pt]{Germanium } \\
\hline
\raisebox{0pt}[13pt][7pt]{MILAN}  &  \raisebox{0pt}[13pt][7pt]{Gran Sasso, Italy}  & \raisebox{0pt}[13pt][7pt]{TeO$_2$ } \\
\hline
\raisebox{0pt}[13pt][7pt]{ORPHEUS}  & \raisebox{0pt}[13pt][7pt]{Bern, Switzerland}  & \raisebox{0pt}[13pt][7pt]{Superconducting Sn grains} \\
\hline
\raisebox{0pt}[13pt][7pt]{ROSEBUD}  &  \raisebox{0pt}[13pt][7pt]{Canfranc, Spain}  &  \raisebox{0pt}[13pt][7pt]{Sapphire } \\
\hline
\raisebox{0pt}[13pt][7pt]{TOKYO}  & \raisebox{0pt}[13pt][7pt]{Nokogiri-yama, Japan} & \raisebox{0pt}[13pt][7pt]{LiF } \\
\hline   
\end{tabular}
\end{center}
\end{table}

To be screened from cosmic ray
background WIMP detectors are located in deep underground laboratories. 
In addition they need to be shielded locally against radioactivity
from surrounding rocks and materials. The
shielding as well as the detector itself have to be fabricated from radiopoor
materials. Background suppression by employing only passive shielding in form of 
radiopoor lead,
copper and other materials is expensive and limited in its effectiveness .
However, large factors in the detection sensitivity can be gained by
using in addition active background recognition methods. Several dark matter 
detectors make
use of this by discriminating  between electron recoils 
(Compton scattering) and nuclear
recoils. In this report only a few dark matter detectors are described.

The CDMS collaboration\cite{nam,gaits,hell} employs two different techniques to measure the energy loss
of a particle in the absorber. The first one utilizes NTD germanium thermistors
bonded to Ge crystals. They call this type of detector BLIP (Berkeley Large
Ionization and Phonon based detector). The present detector consists of 3 Ge
crystals with a weight of 165g each. Each crystal has two thermistor  sensors in
order to be able to crosscheck the signals rejecting direct particle interactions
in one of the thermistors which would yield only small ionisation signals. The
detector is operated at a temperature of 20 mK. The event signals are rather
slow with risetimes of a few ms and fall times of about 20ms.

The second type detector is called ZIP(Z-sensitive Ionization and Phonon-based
detector). It utilizes tungsten aluminum QET¹s (Quasiparticle trapping assisted
Electrothermal feedback Transition edge sensors). This type of sensor covers a
large area of the silicon absorber (100g) with aluminum phonon collector pads,
where phonons are absorbed by breaking Cooper pairs and forming quasiparticles.
The quasiparticles are trapped into a meander of tungsten strips which are used
as transition edge sensors. The release of the quasiparticle energy in the
tungsten strips increases their resistance, which will be observed as a current
change in L detected with a SQUID as indicated in Fig.5. The transition edge
device is voltage biased to take advantage of the electrothermal feedback. The
signal pulses of the ZIP detector have rise times of a few $\mu$s and fall times of
about 50$\mu$s.
 They are much faster than the signals of the BLIP detector since the
ZIP detectors are sensitive to the more energetic nonthermal phonons. Their sensitivity to nonthermal phonons and the pad structure of
the sensors at the surface of the crystal allows for a localization of the event
in the x-y plane (Fig.5).

\begin{figure}[t]
\epsfxsize=20pc 
\epsfysize=15pc 
\hspace*{1.5cm}
\epsfbox{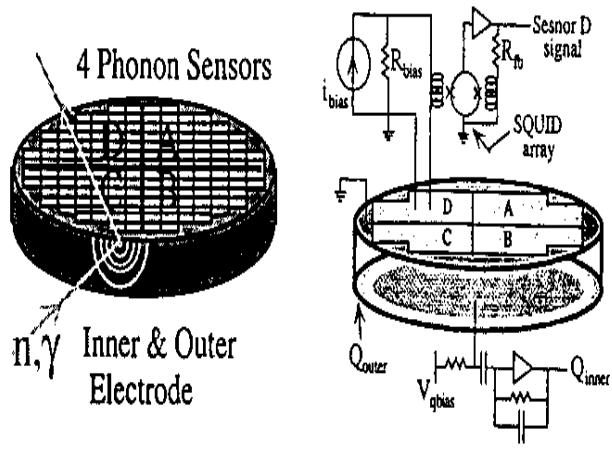} 
\caption{
The schematic of the electrical circuits for
reading out the phonon and charge
signals from a ZIP detector.
 \label{fig:figure5}}
\end{figure}

The energy of the recoiling nucleus or electron in the absorber appears in form
of phonons and electron-hole pairs. A simultaneous measurement of both, the
phonons and electron-hole pairs, in each event allows to discriminate the
nuclear recoils from the electron recoil backround events (Fig.6).This is possible since
for a given deposited  energy the ionization generated by nuclear recoils is
smaller than by electrons. By this method 99$\%$
of electron recoils with energies above a 15 keV can be recognized.

\begin{figure}[t]
\epsfxsize=20pc 
\epsfysize=20pc 
\hspace*{1.5cm}
\epsfbox{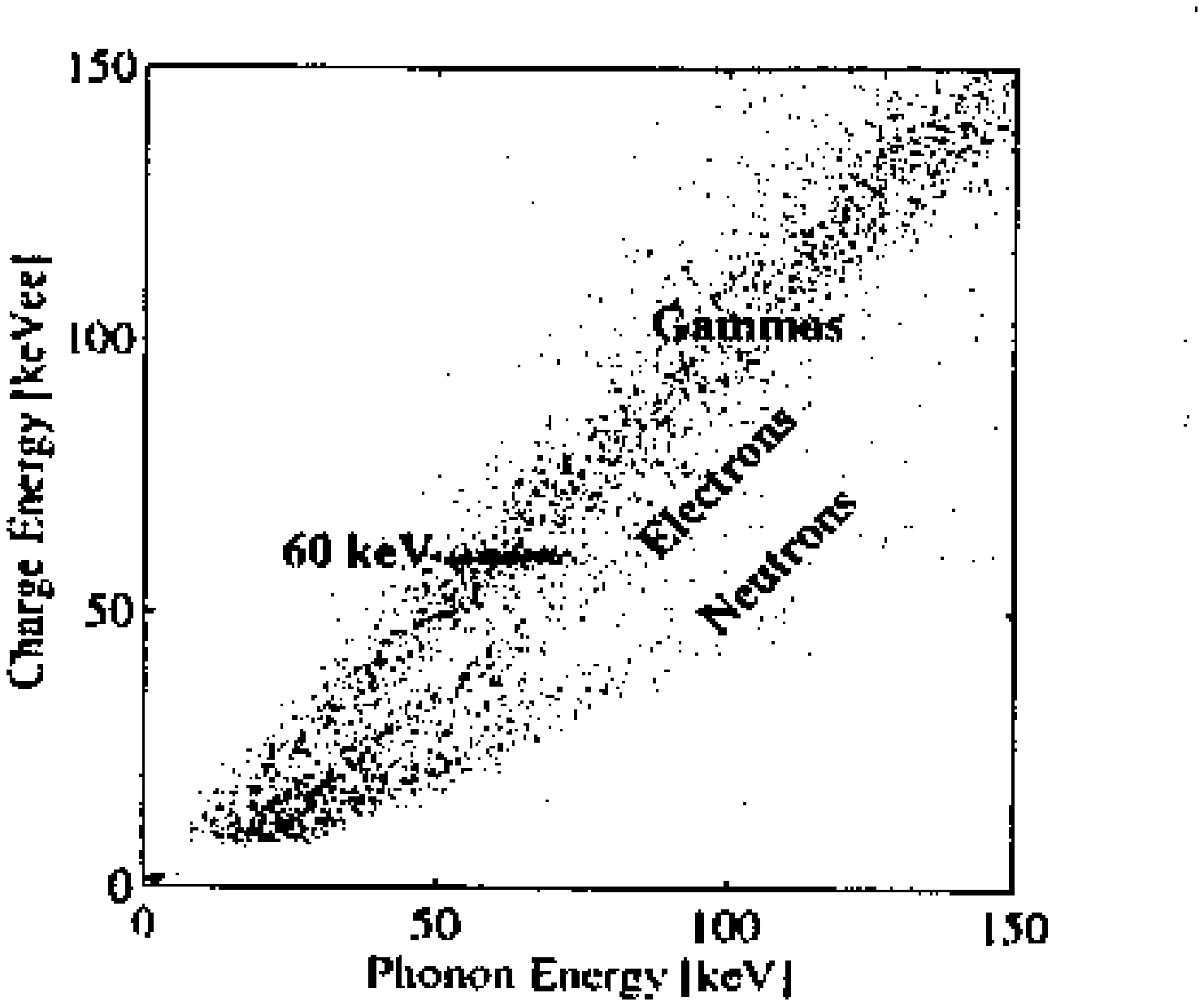} 
\caption{
The charge and phonon energy of $^{14}$C electrons,
$^{241}$Am photons (60 keV) and $^{252}$Cf neutrons
are shown. 
  \label{fig:figure6}}
\end{figure}

The attractive nuclear and electron recoil discrimination applied with the CDMS BLIP 
detector is also realized in the EDELWEISS experiment, which is presently taking data 
in the Frejus tunnel at 4800 m.w.e.\cite{debe}. The detector uses  70g sapphire crystals. 
The collaboration plans to increase the detector mass to about 10kg in the future.

The CRESST collaboration\cite{bra} is using in the first phase of their dark matter
search four 262g sapphire detectors, each equipped with a tungsten TES and electrothermal feedback.
  The detector is set up in the Gran Sasso Laboratory and operates at a temperature of about 15 mK. The pulse height spectrum
measured with a 262g sapphire detector exposed to a x-ray fluorescence source is
shown in Fig.7. The large background towards lower energies is attributed to a
damaged thin Al sheet meant to absorb Auger electrons from the source. The detector
is capable of detecting recoil energies down to about 200 eV.

\begin{figure}[t]
\epsfxsize=20pc 
\epsfysize=20pc 
\hspace*{1.5cm}
\epsfbox{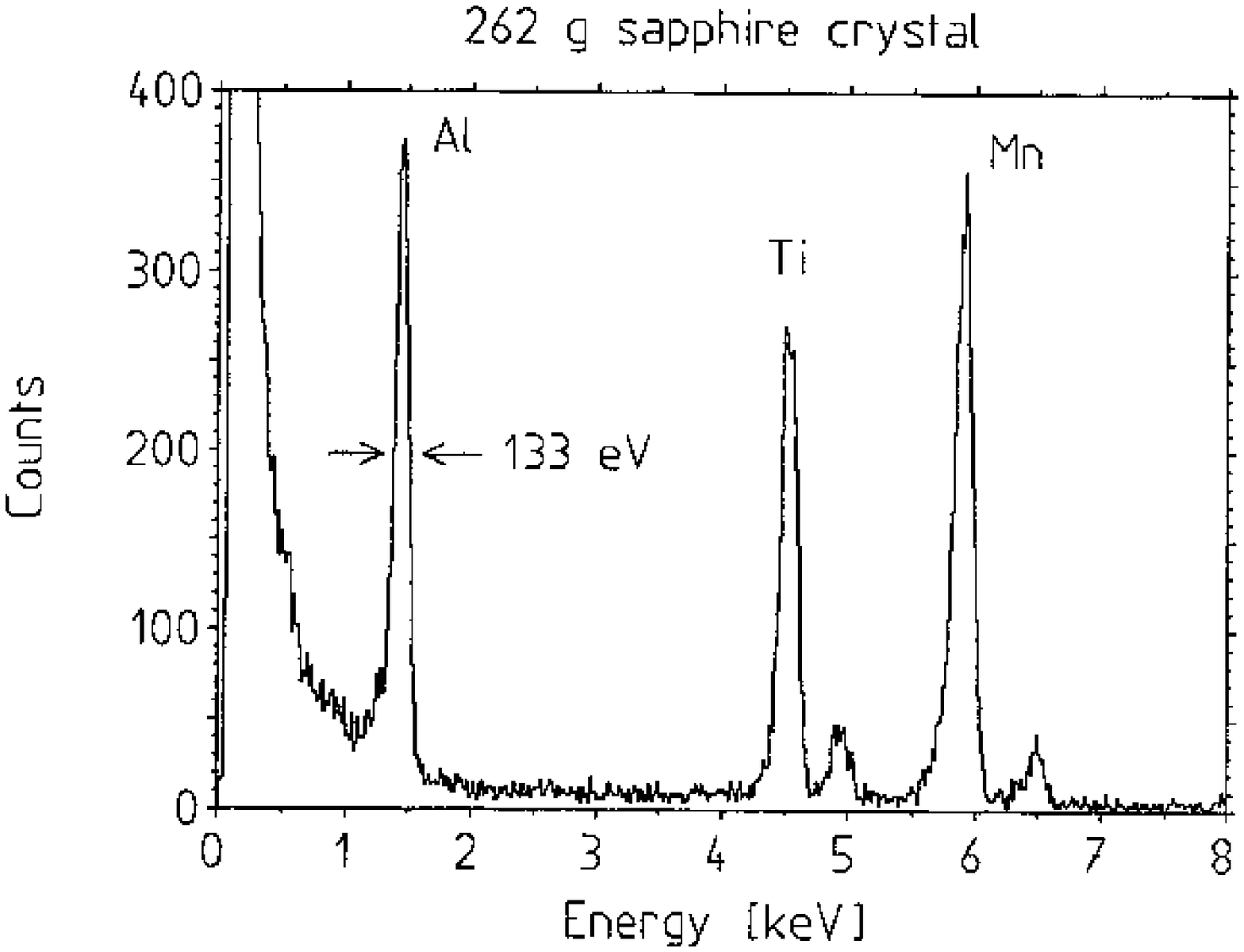} 
\caption{
The pulseheight spectrum measured with a 262 g sapphire detector
exposed to a X-ray fluorescence source is shown. 
\label{fig:figure7}}
\end{figure}

For the future, the CRESST collaboration\cite{bra} is proposing to use CaWO$_4$ (Calcium
tungstate) crystals as scintillating absorbers. By simultaneously measuring the
phonons and the light, they are  able to discriminate between nuclear and 
electron recoils. First results
obtained  with a 6g CaWO$_4$ crystal are shown in Fig.8. The left plot in Fig.8
shows a scatterplot of the energy equivalent of the pulse heigths measured in the
light detector versus those measured in the phonon detector. This scatterplot
was obtained with an electron and a photon source. The right hand plot shows an
additional irradiation with neutrons from an Americium-Beryllium source clearly
demonstrating the line coming from neutron induced nuclear recoils. Above an energy of 
15 keV  99.7$\%$ of the electron recoils can be recognized. At present the CRESST
collaboration is investigating also other suitable scintillating materials. With this method
a powerful tool of active background rejection can be realized, enhancing the WIMP
detection sensitivity of these devices considerably.

\begin{figure}[t]
\epsfxsize=20pc 
\epsfysize=20pc 
\hspace*{1.5cm}
\epsfbox{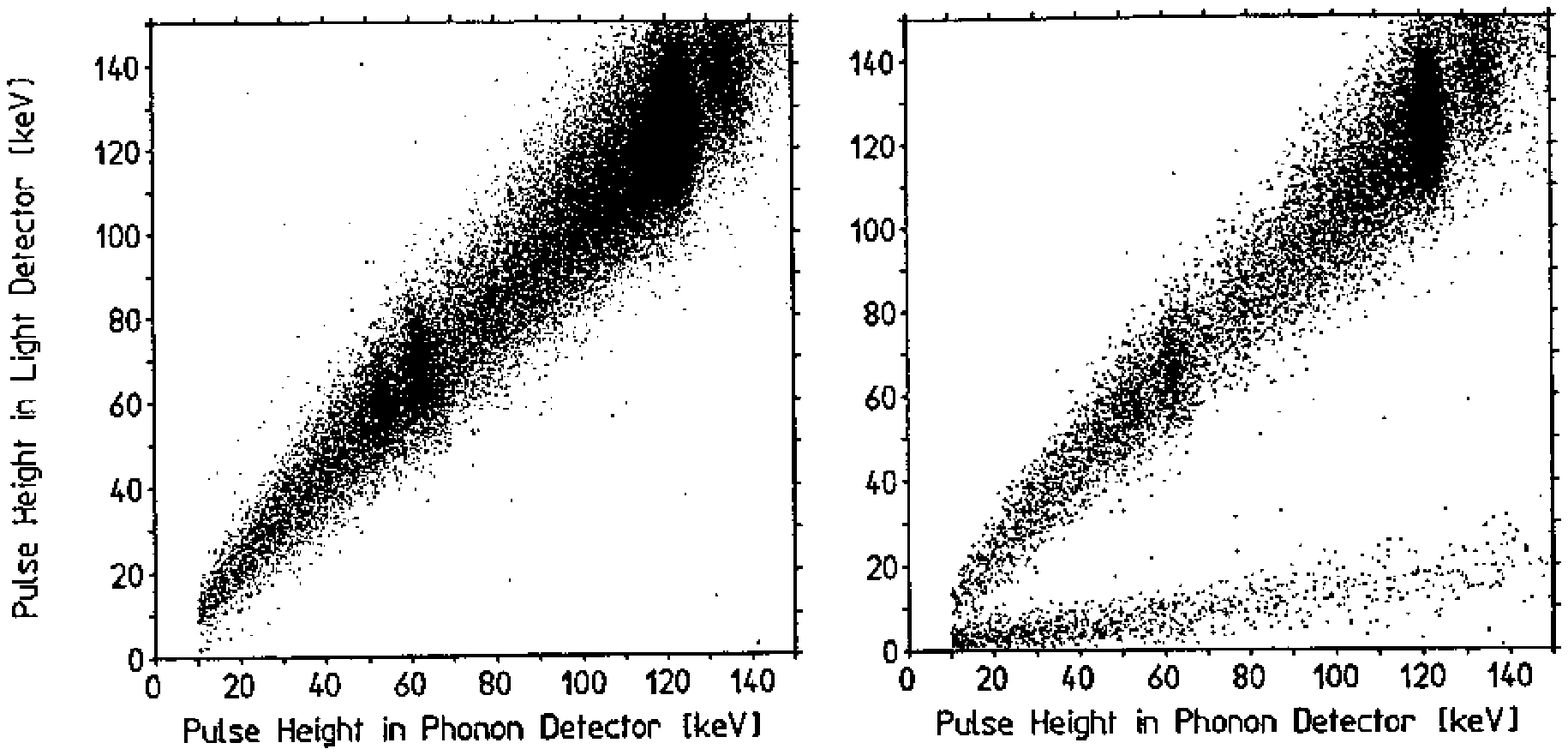} 
\caption{
Left, a scatterplot of the energy equivalent of the pulse
heights measured in the light detector
versus those in the phonon detector.
This was obtained with an electron and photon source.
Right,
the scatterplot with additional irradiation with neutrons from an 
Americium-Beryllium source is shown,
clearly demonstrating the lower
line coming from neutron induced nuclear recoils.
\label{fig:figure8}}
\end{figure}

The ROSEBUD dark matter experiment consists of two 25g and one 50g sapphire absorbers 
which are equipped with NTD Ge thermistors. The experiment has recently been 
set up in 
the Canfranc Underground Laboratory (Spain) at 2450 m.w.e. and started data taking\cite{ce}.

The ORPHEUS collaboration\cite{apbl} developed a SSG detector for a dark matter search. Nuclear recoil measurements with granules made from 
type-1 superconductors (Al,Sn,Zn) have been performed  in a neutron beam at 
the Paul Scherrer Institut  in Villigen\cite{ap}. All these materials turned out to be 
suitable for a dark matter detector. At present, the ORPHEUS collaboration is setting up 
a 1kg SSG detector, made from 30$\mu$m diameter Sn granules, in the Bern Underground 
Laboratory at 70 m.w.e. Among other interesting features, as already mentioned above, 
SSG allows to distinguish ionizing particles which cause many grains to flip from WIMP 
interactions which cause only one grain to flip. This feature will be used to reduce background 
from cosmic rays,
Compton scatterings and radioactive materials. Fig.9 shows a pulseheight 
spectrum obtained with a Zn SSG detector exposed to a neutron beam at the Paul Scherrer 
Institut. The one grain flips are due to neutron scatterings, while the flips of several 
granules are due to Compton electrons also present in the beam.

\begin{figure}[t]
\epsfxsize=20pc 
\epsfysize=20pc 
\hspace*{1.5cm}
\epsfbox{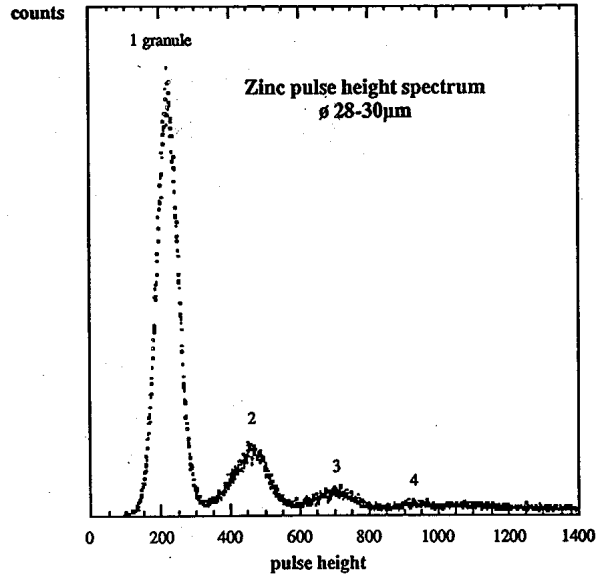} 
\caption{
The pulseheight spectrum of a Zn SSG detector exposed to a beam of neutrons
and Compton electrons is shown.
\label{fig:radish}}
\end{figure}

\section{Double Beta Decay Experiments}

The largest low temperature detector in operation so far is the one of the Milan group 
in the Gran Sasso Laboratory. The detector consists of an array of 20 TeO$_2$ crystals of
340g each, totalling a mass of almost 7kg \cite{ale}. Attached to each crystal are Ge
NTD thermistor thermometers. The detector is mainly used to search for the neutrinoless 
double beta decay of $^{130}$Te, taking advantage of the large isotopic abundance (34$\%$) and
the large transition energy of 2528 keV . Low temperature devices are
particularly well suited for this type of research since they provide energy resolutions
better than or comparable with semiconductors and they can be made of a wide choice of
double beta active nuclei. The TeO$_2$ detector of the Milan group is also used for the
search of WIMPs. However, its recoil energy threshold of 13keV is rather high compared 
to other cryogenic detectors.

For the future E.Fiorini \cite{fio} proposes a large Cryogenic Underground Observatory
for Rare Events (CUORE) in the Gran Sasso Laboratory. The proposal foresees the 
construction of a cryogenic detector containing many such crystals 
(TeO$_2$  or others), totalling a mass of about one ton. From this device very
important new results on double beta decay, WIMP search, solar axions and neutrino 
interactions from artificial sources are expected.

\section{Beta Decay and the Mass of the Electron Neutrino}

Most of the experiments to search for a non-zero mass of the neutrino in beta decay
are using the tritium decay into $^{3}$He, since this transition has a rather low endpoint
energy of 18.6keV. However, so far no experiment was able to establish a
finite mass of the neutrino. Only upper limits were obtained, of which the lowest is
4.35eV \cite{bel}. One of the problems with experiments based on a spectroscopic
measurement of the emitted electron is that they all yield negative values for the square
of the neutrino mass when fitting the electron energy spectrum. This could possibly
be due to final state interactions (like tritium decays into excited atomic levels of 
$^{3}$He),
which lead to a deviation of the expected energy spectrum of the electron. Low temperature
calorimeters reaching similar energy resolutions could perhaps overcome these
difficulties, since they measure the total energy including final state
interactions such as the de-excitation energy of excited atomic levels.

The Genova group has studied the beta decay of $^{187}$Re with a cryogenic microcalorimeter
\cite{fon}.The detector is a rhenium single crystal (2mg) coupled
to a Ge NTD thermistor. Rhenium is a superconductor with a critical temperature of 1.7 K. 
Natural Re contains 62.$\%$ of
$^{187}$Re with an endpoint energy of about 2.6 keV. The operating temperature of 
the detector was T = 90 mK. In their
first attempt they made a precise measurement of the endpoint energy of (2.64$\pm $0.02)keV
and of the half-life of (62$\pm$6) x $10^{9}$ years.  The Milan group has also started to 
develop
thermal detectors for a $^{187}$Re neutrino mass experiment \cite{ales}. With cryogenic 
calorimeters a higher sensitivity to low neutrino masses may be achievable in the future,
provided that an energy resolution of less than 10 eV can be obtained and the statistics
at the endpoint energy can be improved. The latter may raise a problem for thermal phonon
detectors, since their signals are rather slow
and therefore limit the counting rate capability considerably. This problem may be 
overcome by a new approach reported at this conference \cite{gom}, 
which uses the fast transition of Re from a metastable superconducting into a 
normal state.

With their Re cryogenic microcalorimeters the Genova group was able to detect interactions
between the emitted beta particle and its local environment, known as beta environmental
fine structure \cite{ga}. Their encouraging results show that cryogenic microcalorimeters
may also offer a new way to study molecular and crystalline structures.

\section{Single Photon Counting Cryogenic Detectors for Applications in Astrophysics}

A detector capable of measuring individual photons with high efficiency
for a wide range of wavelengths (optical to near infrared) and at the same time
the arrival time of every single photon falling upon the detector has  sofar not
existed. However, as recently demonstrated by the ESTEC (European Space Agency) group,
superconducting tunnel junctions can be used as single photon counting spectroscopic
detectors \cite{pea}. The detection principle is based on the fact that for a 
superconductor with a $\approx$1meV gap energy  an optical photon of $\approx$1eV
represents a large amount of energy. Thus a photon impingeing on a superconductor, 
like for example
tantalum, can create many quasiparticles leading to a measurable tunnel current across a
voltage biased junction. The tantalum STJ developed by the ESTEC group is shown in Fig.10.
The device consists of a 20 x 20$\mu $m$^{2}$ and 100nm thick epitaxial tantalum film on a sapphire
substrate with a 30nm thick aluminum trapping layer on top. The tunnel barrier is 1 nm
thick and consists of oxidized aluminum. Quasiparticles produced after photoabsorption in
the tantalum layer will be trapped by the aluminum film close to the barrier. The initial
fluctuation in the number of charge carriers created by photoabsorption combined with the
tunnel noise leads to an overall limiting resolution of a junction in terms of wavelength.
In Fig.11 the resolution obtained with tantalum and niobium based devices are shown 
\cite{po}. Also shown  is the expected performance of molybdenum and hafnium STJs, which are also under development by the ESTEC
group. The obtained quantum efficiency for the tantalum STJ is 70$\% $
for wavelengths of 200-600nm with a cutoff below 200 nm due to the sapphire substrate. It 
is possible to extend the short wavelength limit to 110 nm by replacing the sapphire substrate with
magnesium fluoride.  The reflectivity of the sapphire substrate and the tantalum reduces 
the quantum efficiency in the infrared considerably, a feature which can be used as an 
infrared filter for some applications.

\begin{figure}[t]
\epsfxsize=20pc 
\epsfysize=20pc 
\hspace*{1.5cm}
\epsfbox{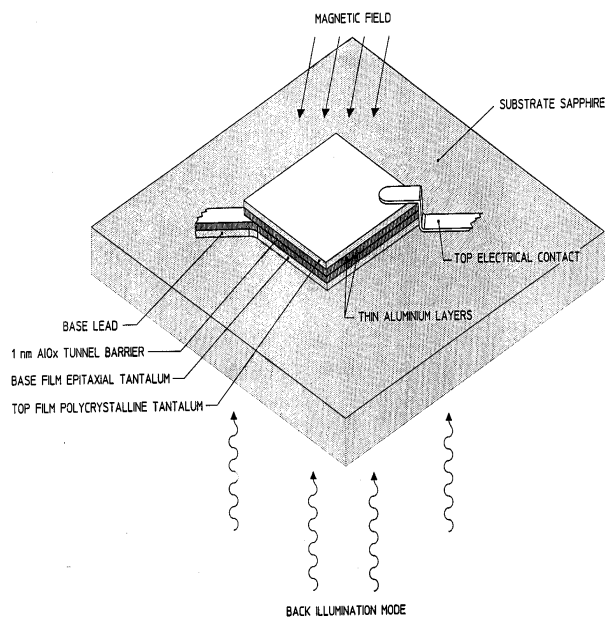} 
\caption{
The schematic of the tantalum tunnel junction is shown.
The orientation of the magnetic field
(parallel to the junction) to suppress the Josephson
current is indicated.
 \label{fig:figure10}}
\end{figure}

\begin{figure}[t]
\epsfxsize=35pc 
\epsfysize=25pc 
\hspace*{-1.5cm}
\epsfbox{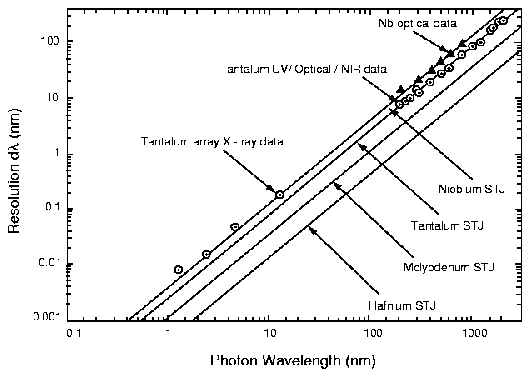} 
\caption{
The resolution of tunnel junctions made from superconductors
as a function of wavelength is shown. 
 \label{fig:figure11}}
\end{figure}

A cryogenic camera consisting of STJ pixel arrays is an attractive device for astronomical 
observations . The single photon detection capability allows to observe very faint
and distant objects and to determine their distance through red shift. A cryogenic camera
with a 6x6 array of tantalum STJ's (with a pixel size of 25x25$\mu $m$^{2}$) has been recently
(Feb.1999) installed in the William Herschel Telescope on La Palma. For a first proof of
principle of this novel technique, the telescope was directed towards the Crab pulsar, whose 
already known periodicity of 33ms was measured by the cryogenic camera as shown in 
Fig.12 \cite{pe}. The
photon arrival time information was recorded with an accuracy of about ±5$\mu $s with respect
to GPS timing signals. The results nicely demonstrate the capabilities of this technology
for future astrophysical observations.

\begin{figure}[t]
\epsfxsize=20pc 
\epsfysize=20pc 
\hspace*{1.5cm}
\epsfbox{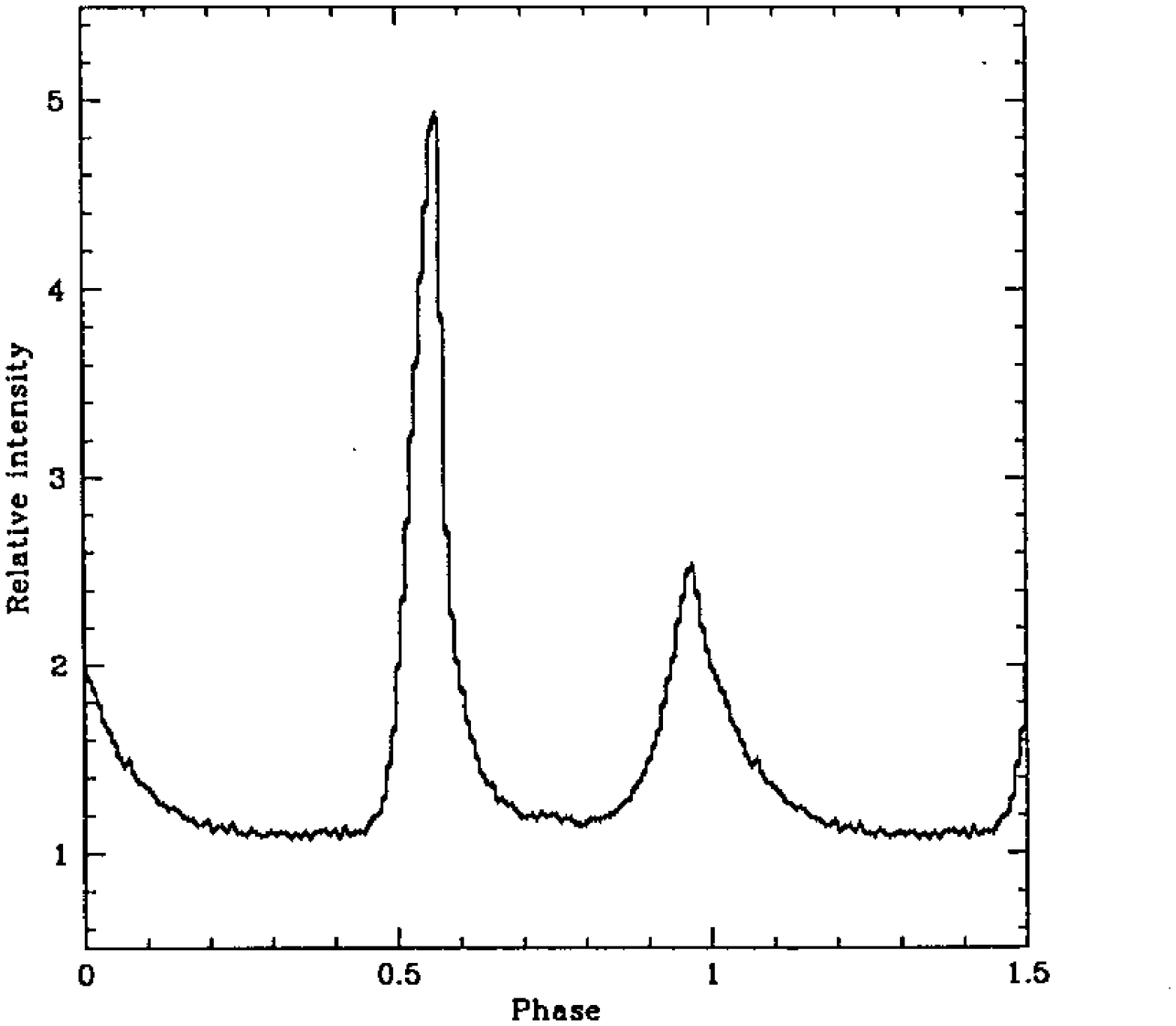} 
\caption{
The pulse profile for the Crab pulsar from the 6. Febr. 1999 data
over the wavelength range 310-610 nm (128 phase bins) is shown.
 \label{fig:figure12}}
\end{figure}

\section{Microcalorimeters for X-ray Astronomy}

Cryogenic microcalorimeters provide major improvements over conventional detection systems 
in x-ray astronomy. For example gratings and Bragg crystal spectrometers have an excellent 
energy resolution, but at the expense of a very small acceptance and energy range. Scanning 
over an entire x-ray spectrum (0.05 to 1keV) of an extended source requires  very long and 
expensive observation times . On the other hand solid state detectors with 
large angular acceptance do not reach energy resolutions better than 100eV. In 
contrast cryogenic microcalorimeters can  measure the energy of each incident x-ray,
simultaneously covering the entire energy range of interest. In addition they have no 
limitation on the acceptance angle. However the deployment of cryogenic
detectors for experiments in space poses a new technical challenge. Pioneering work in this 
direction has been done by the Wisconsin, Maryland, NASA Goddard  (WMN) group \cite{de,kil}.

The WMN group has developed a microcalorimeter array consisting of 36 0.5mmx2mm x-ray
detectors with ion implanted silicon thermistors and HgTe absorbers. The microcalorimeters are operated at 60mK. The cryostat consists of an 
adiabatic demagnetization refrigerator inside a four liter liquid helium dewar.The  
cryogenics together with the detector fit into a cylinder of about 35 cm diameter and 
40 cm length. The detector was flown on a suborbital sounding rocket to observe the 
interstellar diffuse  x-ray  background in the range of 0.05 to 1keV. The sounding rocket
reached an altitude of about 235km. The flight provided 240s of observation time 
above 165km, enough to make  
reasonable spectroscopic measurements. During observation the Helium bath temperature
was regulated to better than 1$\mu $K. The energy resolution obtained on single pixels
ranges from 3.8 - 4.8eV FWHM \cite{camm}. When recovering the payload after the flight the 
dewar contained still some liquid helium.

The physics of the diffuse interstellar x ray background is not very well understood. It seems
that a large component is due to collisional excitations of particles in an interstellar gas 
with temperatures of a few $10^{6}$K. A detailed spectral analysis would allow to determine
the physical state and the composition of the gas. Since the interstellar gas occupies a large 
fraction of the volume within the galactic disc, it plays a major role in the formation of stars 
and the evolution of the galaxy. By employing a conical foil mirror in front of the microcalorimeters 
the WMN group is also planning to perform spectroscopic observations of supernova remnants in 
the near future. 

\section{Transition Edge Microcalorimeters for Industrial Applications}

A group of the National Institute of Standards and Technology (NIST) in Boulder, Colorado
(USA) has developed  high performance x-ray microcalorimeters based on superconducting
transition edge sensors (TES) \cite{lad}. The detectors are cooled to 70 mK by a compact adiabatic 
demagnetization refrigerator and can be mounted on a scanning electron microscope. These 
devices have features which make them very attractive for industrial applications in x-ray 
microanalysis. For example the excellent energy resolution of these devices enables to measure
the chemical shifts in the x-ray spectra caused by changes in the electron binding due to 
chemical bonding. Another application is the analysis of contaminant particles and defects
for the semiconductor industry. This technology looks also promising to be used in detecting
accelerated large masses such as proteins and DNA in a spectrometer. This 
method would  allow to  make DNA sequencing several orders of magnitude faster than 
current gelelectrophoresis methods \cite{twere,hilt}.

The main advantage of microcalorimeters as compared to more conventional x-ray detection systems
(Bragg spectrometers and solidstate detectors) is their ability to cover the entire
x-ray spectrum of interest simultaneously and to resolve the closely spaced lines with high
resolution. However, microcalorimeters are intrinsically slow devices with limited counting 
rate capabilities and with small effective surfaces. To cope with these problems the NIST 
group has developed the following improvements. They are using TES sensors
with an electrothermal feedback system which allows to decrease the time it takes for the microcalorimeter
to reach thermal equilibrium. This way a gain in counting rate of up to a factor 100 could be achieved.
To compensate for the small detector surface they developed an x-ray focussing 
device using polycapillary optics. The device consists of many fused tapered glass capillaries
which focus the x-rays by means of internal reflection onto the microcalorimeter increasing its 
effective area. The excellent performance of the NIST microcalorimeter as compared to a Si(Li) solid state device 
is clearly demonstrated in Fig.13. The microcalorimeter has an energy resolution of 3eV, an 
effective surface area of 4mm$^{2}$ and a count rate capability of 1000 per second. A cross sectional view of a TES microcalorimeter developed
by the NIST group is shown in Fig.14. Currently the NIST researchers are working on further improving
the energy resolution and the counting rate capability of their calorimeters. Nevertheless 
they already demonstrated the quality of these devices for industrial and other applications.

\begin{figure}[t]
\epsfxsize=20pc 
\epsfysize=20pc 
\hspace*{1.5cm}
\epsfbox{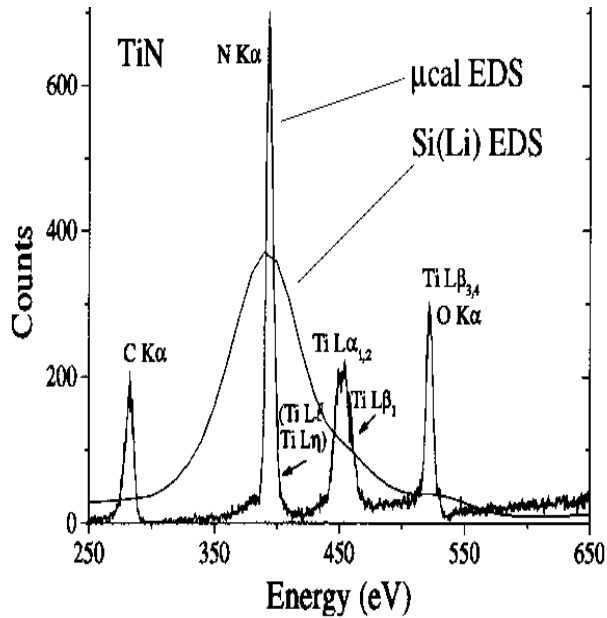} 
\caption{
The energy resolution of a Si(Li)
solid state device versus a energy dispersive (EDS) micro-calorimeter
is shown on the example of TiN, where
TiN is an important interconnect and diffusion barrier used in the
semiconductor industry. 
  \label{fig:figure14}}
\end{figure}

\begin{figure}[t]
\epsfxsize=20pc 
\epsfysize=20pc 
\hspace*{1.5cm}
\epsfbox{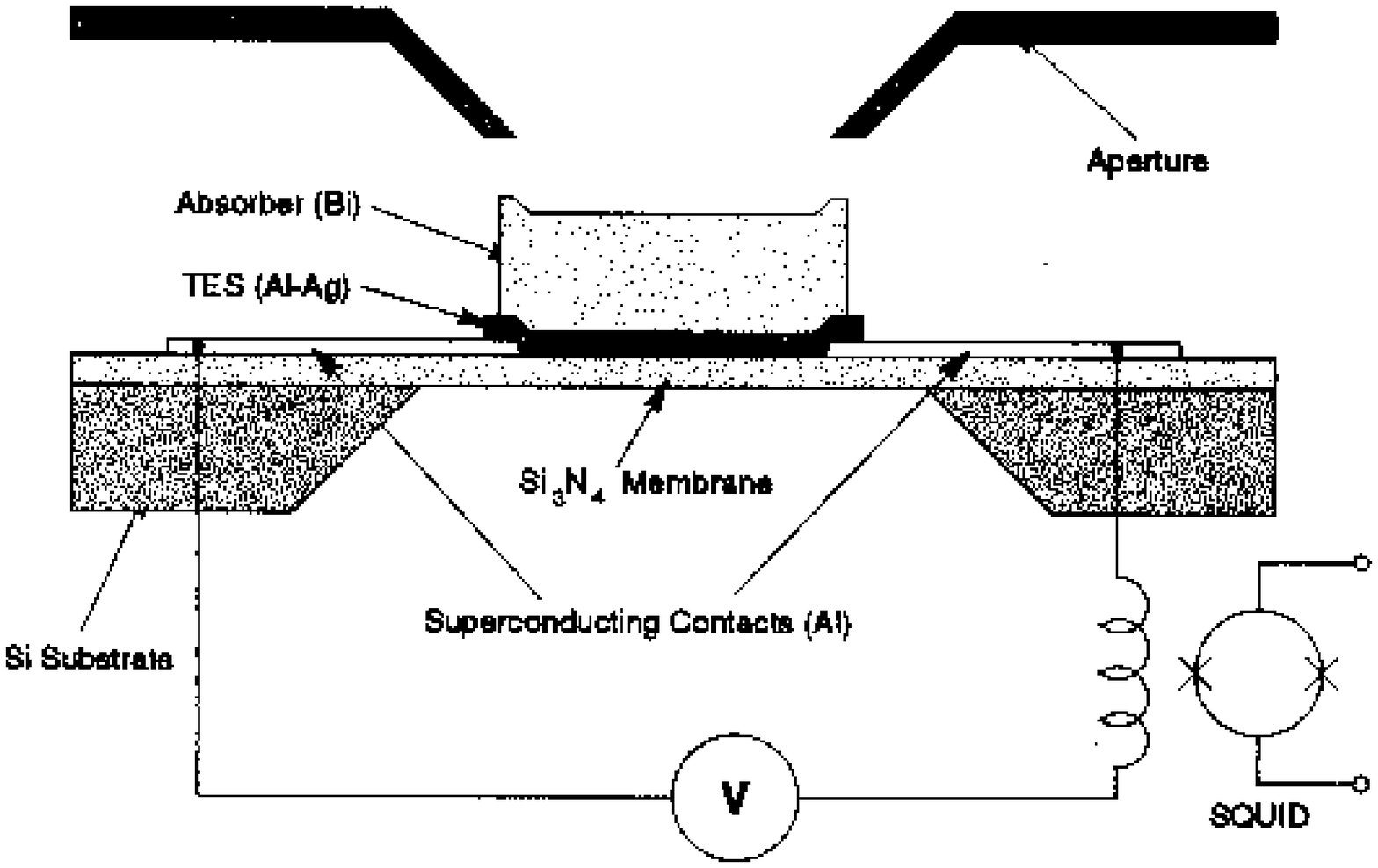} 
\caption{
The cross-sectional view of the TES micro-calorimeter is shown.
The X-rays are passing through an aperture and are absorbed in the
Bi film.
  \label{fig:figure13}}
\end{figure}

\section{Conclusions}

Cryogenic detectors allow to explore new domains in physics and astrophysics which have
sofar not been so easily accessible with other devices. 
The physics of cryogenic detectors, however,
is more complicated than that of ionization devices, therefore rather long research and development
times are required. The cryogenic technology is very advanced and commercially available.
It allows to perform experiments underground and in future even in orbit. Massive 
(several kg) cryogenic detectors for dark matter and double beta decay experiments are 
already in operation. For the future detectors with 100 kg mass and beyond are planned. 
The application of cryogenic detectors in other fields of research as well as in industry has 
been growing rapidly in the last years.

Enjoying the warm and sunny weather in Lisbon  one easily forgets
that there are things which work better in the cold.

 \section*{Acknowledgments}
 I would like to express my very special gratitude to Amelia Maio and 
 the organizers of this conference for the excellent work they have done
 to make this conference a great success. My warmest thanks also to S.Janos 
 for the many enlightening discussions during the preparation of this talk
 and for his careful reading of this manuscript. Furthermore I would like to thank
 R.Cristiano, E.Fiorini, R.Gaitskell, G.Hilton, H.Hoevers, D.Mac Cammon, T.Peacock, F.Proebst, 
 B.Saduolet and D.Wollman for providing me with material for this talk.
 This work was supported by the Swiss National Science Foundation.


\begin{thebibliography}{99}

\bibitem{si}Simon F., Nature 135 (1935) 763
\bibitem{an}Andrews D.H. et al., Phys.Rev. 76 (1949) 154
\bibitem{wo}Wood G.H., White B.L., Appl.Phys.Lett. 15 (1969) 237 and 
Can.J.Phys. 51 (1973) 2032
\bibitem{be}Bernas H. et al., Phys.Lett. A 24 (1967) 721
\bibitem{dr}Drukier A. and Vallette C., NIM 105 (1972) 285
\bibitem{ni}Niinikoski T.O., Udo F. CERN NP Report 74-6 (1974)
\bibitem{fi}Fiorini E., Niinikoski T.O., NIM 224 (1984) 83
\bibitem{dru}Drukier A., Stodolsky L., Phys.Rev. D 30 N11 (1984) 2295
\bibitem{pre}Pretzl K., Schmitz N., Stodolsky L., eds. Low-Temperature Detectors for 
Neutrinos and Dark Matter. Berlin: Springer-Verlag (1987)
\bibitem{go}Gonzalez-Mestres L., Perret-Gallix D., eds., Low-Temperature Detectors for
Neutrinos and Dark Matter II. Gif-sur-Yvette, France: Ed. Frontieres (1988)
\bibitem{br}Brogiato L., Camin D.V., Fiorini E., eds., Low-Temperature Detectors for
Neutrinos and Dark Matter III. Gif-sur-Yvette, France: Ed. Frontieres (1989)
\bibitem{bo}Booth N.E., Salmon G.L., eds., Low-Temperature Detectors for Neutrinos and 
Dark Matter IV. Gif-sur-Yvette, France: Ed. Frontieres (1992)
\bibitem{la}Labov S.E., Young B.A., eds., Proc. 5th Int. Workshop on Low Temp. 
Detectors LTD5, Berkeley, CA. J.Low-Temp.Phys. 93(3/4) (1993) 185-858
\bibitem{ot}Ott H.R., Zehnder A., eds., Proc. 6th Int. Workshop on Low Temp. Detectors 
LTD6, Beatenberg/Interlaken, Switzerland. Nucl.Instrum.Methods A 370 (1996) 1-286
\bibitem{co}Cooper S. ed., Proc. 7th Int. Workshop on Low Temp. Detectors LTD7,
Munich, Germany
\bibitem{ba}Barone A. ed., Proc. Superconductive Particle Detectors, Torino 26-29 Oct 
1987, World Scientific
\bibitem{bar}Barone A., Nucl.Phys. B (Proc.Suppl.) 44 (1995) 645;
\bibitem{tw}Twerenbold D., Reports on Progress in Physics 59 (1996) 349;
\bibitem{kra}Kraus H., Supercond.Sci.Technol. 9 (1996) 827;
\bibitem{boot}Booth N., Cabrera B. and Fiorini E. Annu.Rev.Nucl.Part.Sci. 46 (1996) 471
\bibitem{al}Alessandrello A. et al., Phys.Rev.Lett. 82 (1999) 513
\bibitem{irw}Irwin K.D., Appl.Phys.Lett. 66 (1995) 1998
\bibitem{baro}Barone A. and Paterno G., 1984 Physics and Applications of the Josephson 
Effect (New York: Wiley)
\bibitem{ka}Kaplan B. et al., Phys.Rev. B 14 (1976) 4854 and Erratum Phys.Rev. B 15
(1977) 3567
\bibitem{ale}Alessandrello A. et al., Phys.Rev. B ... (in press)
\bibitem{fio}Fiorini E., Physics Reports 307 (1998) 309
\bibitem{bel}Belesev A. et al., Phys.Lett. B 350 (1995) 263
\bibitem{fon}Fontanelli F. et al. NIM A 370 (1996) 247
\bibitem{ales}Alessandrello A. et al., Nucl.Phys. B (Proc.Suppl.) 70 (1999) 230-232
\bibitem{gom}Gomes M.R. et al.,see Proceedings of this Conference 
\bibitem{ap}Abplanalp M. et al., NIM A 360 (1995) 616
\bibitem{ga}Gatti F. et al., Nature Vol.397 (1999) 137
\bibitem{pea}Peacock T. et al., Astron.Astrophys.Suppl.Ser. 127 (1998) 497
\bibitem{po}Poclaert A. et al., SPIE 2808 (1996) 523
\bibitem{pe}Perryman M. et al., ESLAB 1999/021/SA
\bibitem{de}Deiker S. et al., in Proc. 7th Int. Workshop on Low Temp. Detectors LTD7,
Munich, Germany,S.Cooper ed,C4 (1997) 108
\bibitem{camm}Mc Cammon D., private communication
\bibitem{lad}Ladbury R.in Physics Today (July 1998) 19 and references therein
\bibitem{hilt}Hilton G. et al., Nature 391 (1998) 672
\bibitem{hil}Hilton G. private communication
\bibitem{twere}Twerenbold D., NIM A 370 (1996) 253
\bibitem{cris}Cristiano R. et al., Appl.Phys.Lett.74,N 22 (1999)
\bibitem{pr}Pretzl K., Particle World Vol.1, No.6 (1990) 153
\bibitem{pret}Pretzl K., Journ.of Low Temp.Physics, Vol.93,No.3/4 (1993) 439
\bibitem{cal}Calatroni S. et al., to be published in NIM, the Proceedings of LTD-8 (1999)
\bibitem{bra}Bravin M. et al., MPI-PhE/98-22
\bibitem{labov}Labov S.et al., Proceedings LTD-7, editor S.Cooper (1997) 82
\bibitem{peac}Peacock T. et al., Astron.Astrophys.Suppl.Ser.123,(1997) 581
\bibitem{nam}Nam S. et al., Proceedings LTD-7, editor S.Cooper (1997) 217
\bibitem{gaits}Gaitskell R. et al., Proceedings LTD-7, editor S.Cooper (1997) 221
\bibitem{hell}Hellmig J. et al., to be published in NIM, the Proceedings of LTD-8 (1999)
\bibitem{colli}Colling P. et al., NIM A354,(1995) 408
\bibitem{vande}van den Brandt et al, Nucl.Physics B (Proc.Suppl.) 70 (1999) 101
\bibitem{apbl}Abplanalp M. et al., NIM A 370 (1996) 227
\bibitem{debe}De Bellefon A. et al., Astroparticle Physics 6 (1996) 35
\bibitem{ce} Cebrian S. et al., Astroparticle Physics  10 (1999) 361
\bibitem{kil}Kibourne Stahle K. et al., Physics Today August 1999, page 32
\end{thebibliography}
\end{document}